\RequirePackage{fix-cm}
\documentclass[twocolumn]{svjour3}          
\smartqed  
\usepackage{graphicx}
\usepackage{cite}
\usepackage{subcaption}
\usepackage{float} 
\usepackage{amsmath} 
\usepackage{natbib}
\usepackage{breakcites}
\begin{document}
%

\title{Synthesis of Convection Velocity and Turbulence Measurements in Three-Stream Jets}


\author{Marcie Stuber         \and
        K. Todd Lowe \and
        Wing F. Ng 
}


\institute{   Advanced Propulsion and Power Laboratory \\
              Virginia Tech\\
              Blacksburg, VA USA 24061\\
              Tel.: +540-231-7650\\
              Fax: +540-231-9632\\
              \email{kelowe@vt.edu}           
}

\date{Received: date / Accepted: date}

\maketitle

\begin{abstract}
Key flow regions linked to jet noise sources are investigated through comparison of convection velocity and turbulence measurements in high speed three-stream nozzles. A time-resolved Doppler global velocimetry (DGV) instrument was applied in the Nozzle Acoustic Test Rig (NATR) at NASA’s Aero-Acoustic Propulsion Lab to measure 250 kHz repetition laser scattering signals arising from seeding particles at 32 points in high speed flow. Particle image velocimetry (PIV) measurements previously reported by NASA provided mean velocity and turbulence intensity. Results for convection velocity of the particle concentration field were obtained from the DGV data for three-stream nozzle configurations using a cross-correlation approach. The three-stream cases included an axisymmetric and an asymmetric, offset nozzle configuration. For the axisymmetric case, areas of high convection velocity and turbulence intensity were found to occur 4 to 6 area-equivalent nozzle diameters downstream from the nozzle exit. Comparison of convection velocity between the axisymmetric and offset cases show this same region as having the greatest reduction in convection velocity due to the offset, up to 20\% reduction for the offset case compared to the axisymmetric case. These findings suggest this region along the centerline near the end of the potential core is an important area for noise generation with jets and contributes to the noise reductions seen from three-stream offset nozzles. 
\end{abstract}
\section*{Nomenclature}
\begin{table}[H]
\centering
\label{tab:a}
\begin{tabular}{ll}

$a$	 &	Sound speed, m/s\\

$D$	 &	Diameter, m\\

$M$	 &	Mach number\\

$NPR$	 &	Nozzle pressure ratio, $p_0/p_\infty$\\

$NTR$	 &	Nozzle temperature ratio, $T_0/T_\infty$\\

$R$	 &	Correlation function\\

$Re$	 &	Diameter Reynolds number\\

$s$	 &	Signal dependent variable, a.u.\\

$t$	 &	time, s\\

$T$	 &	Sampling period, s\\

$u^\prime$	 &	Root-mean-square turbulent velocity, m/s\\

$U$	 &	Mean stream-wise velocity, m/s\\

$x$	 &	Stream-wise coordinate, m\\

$y$	 &	Radial coordinate, m\\

$\eta$	 &	Radial lag variable, m\\

$\tau$	 &	Temporal lag variable, s\\

$\xi$	 &	Stream-wise lag variable, m\\
&\\
Subscripts: & \\
$b$	& Bypass nozzle conditions\\

$c$	& Eddy convection or core nozzle conditions\\

$eqA$	& Equivalent area\\

$j$	& Core jet exit condition\\

$r$	& Convective ridge locus\\

$t$	& Tertiary nozzle conditions\\

$\infty$		& Ambient condition\\
&\\
Abbreviations: & \\
AAPL	 &	Aero-Acoustic Propulsion Lab\\

DGV	&	Time-resolved Doppler Global Velocimetry\\

HFJER &	High Flow Jet Exit Rig\\

NASA		& National Aeronautics and Space Administration\\

NATR &		Nozzle Acoustic Test Rig\\

PIV	&	Particle image velocimetry\\

PMT	&	Photomultiplier tube\\

TR	&	Time resolved\\

VT	&	Virginia Tech\\
&\\
&\\
&\\
&\\
&\\
&\\
&\\
&\\
\end{tabular}
\end{table}

%
%
%
%
%
%
%
%
%
%
%


\section*{Introduction}
Noise from high speed jet plumes has been a problem in the aviation industry since the introduction of turbine engines (\cite{lilley1996radiated}). Although advances have been made in the reduction of jet noise, these have primarily been for transonic and subsonic commercial applications. Renewed interest in supersonic transport vehicles through the Commercial Supersonic Technology project of the Advanced Air Vehicles Program at NASA has prompted research in mitigating jet noise in heated, high speed jet plumes (\cite{huff2016perceived}).  Based on the set performance and noise goals of the program, new engine technologies are being researched. One of the more promising ideas is a variable cycle or adaptive cycle engine which would offer a third exhaust stream, in addition to the bypass and core streams that are already commonplace in aircraft engines (\cite{huff2016perceived}).  Recent computational and experimental work has shown that nozzles with a tertiary stream offer a potential for noise reduction (\cite{henderson2012aeroacoustics}; \cite{ henderson2015measurements}). Additional studies have shown that asymmetry in jet plumes reduced jet noise on the locally thick side (\cite{papamoschou2014aeroacoustics}; \cite{papamoschou2012modeling}; \cite{papamoschou2017very}). The asymmetry was introduced by offsetting a secondary or tertiary stream in the radial direction.\

Research on heated, high speed jet noise is also of interest for military aircraft. Due to stricter performance and size requirements, significant advances have not been made for jet noise from military tactical engines. The US Department of Veterans Affairs (VA) spends over \$1 billion each year on hearing loss claims, with 28\% of those from the Navy, a branch with high exposure rates to intense jet noise (Naval Research Advisory Committee Report 2009). In the past, to protect service persons from hearing loss due to jet noise, the Navy put significant effort into improving hearing protection for service persons. While advances in hearing protection have been made, with the newest designs offering protection up to 47 dB, noise levels for those working in close proximity to aircraft can exceed 140 dB (\cite{keefe2015magazine}). New efforts have now been made by the Navy to better understand the fundamental mechanisms causing noise in jet plumes at conditions seen in military aircraft engines. \

Overall, a better understanding of the physics producing jet noise can lead to better engine design with reduced jet noise. Further, continued research on techniques to reduce noise, such as the use of a third stream, is vital to address the jet noise problem in both supersonic transport vehicles and military tactical aircraft. \

\cite{papamoschou2014reduction} have shown that a component of high speed jet noise, the radiation efficiency, is a function of the acoustic convection Mach number of the turbulent eddies in the jet. The acoustic eddy convection Mach number is defined as $U_c/a_\infty $, the eddy convection velocity divided by the ambient speed of sound. The radiation efficiency is therefore a function of the eddy convection velocity of the turbulent structures in the jet. Past computational work (\cite{papamoschou2014reduction}) has also shown that radiation efficiency scales non-linearly with the eddy acoustic convection Mach number. Due to the non-linearity, a reduction in the acoustic eddy convection Mach number is the most effective method for reducing jet noise. Measuring convection velocity in jets, therefore, will provide insight into the mechanisms for reducing noise in heated, three-stream axisymmetric and offset jet streams which have been devised based upon the hypothesis that they reduce convection velocity.\

To measure convection velocity, a time-resolved Doppler global velocimetry (DGV) instrument was applied to heated three-stream axisymmetric and asymmetric offset nozzle configurations at the NASA Aero-Acoustic Propulsion Lab (AAPL) located at Glenn Research Center (\cite{ecker2016scale}). The current work aims to better understand the fundamental physics which cause tertiary streams and offset streams to reduce jet noise. It will be shown that comparison of convection velocity and flow characteristics obtained using two component PIV allows identification of key areas in the flow responsible for noise generation. Investigation of differences between convection velocity in the axisymmetric and asymmetric configurations will reveal differences which contribute to the noise reduction in the offset configuration. Defining these areas improves nozzle designs by targeting regions that will be most efficient in reducing jet noise. This work presents the analysis of experimentally measured turbulence characteristics to provide a better understanding of the mechanisms causing noise generation in heated three-stream jets. \

\section*{Experimental Methods}

Measurements are presented from two different flowfield diagnostics, two-component PIV and time-resolved DGV. The two-component PIV was used to measure axial and radial velocity in an effort led by colleagues at the NASA Glenn Research Center. For more details on the PIV setup and instrumentation, the reader is referred to the article by \cite{henderson2016characterization}. Convection velocities were measured using the DGV instrument previously developed (\cite{ecker2014development}), discussed further to follow.\

\begin{figure*}
\centering
  \includegraphics[width=1.00\textwidth]{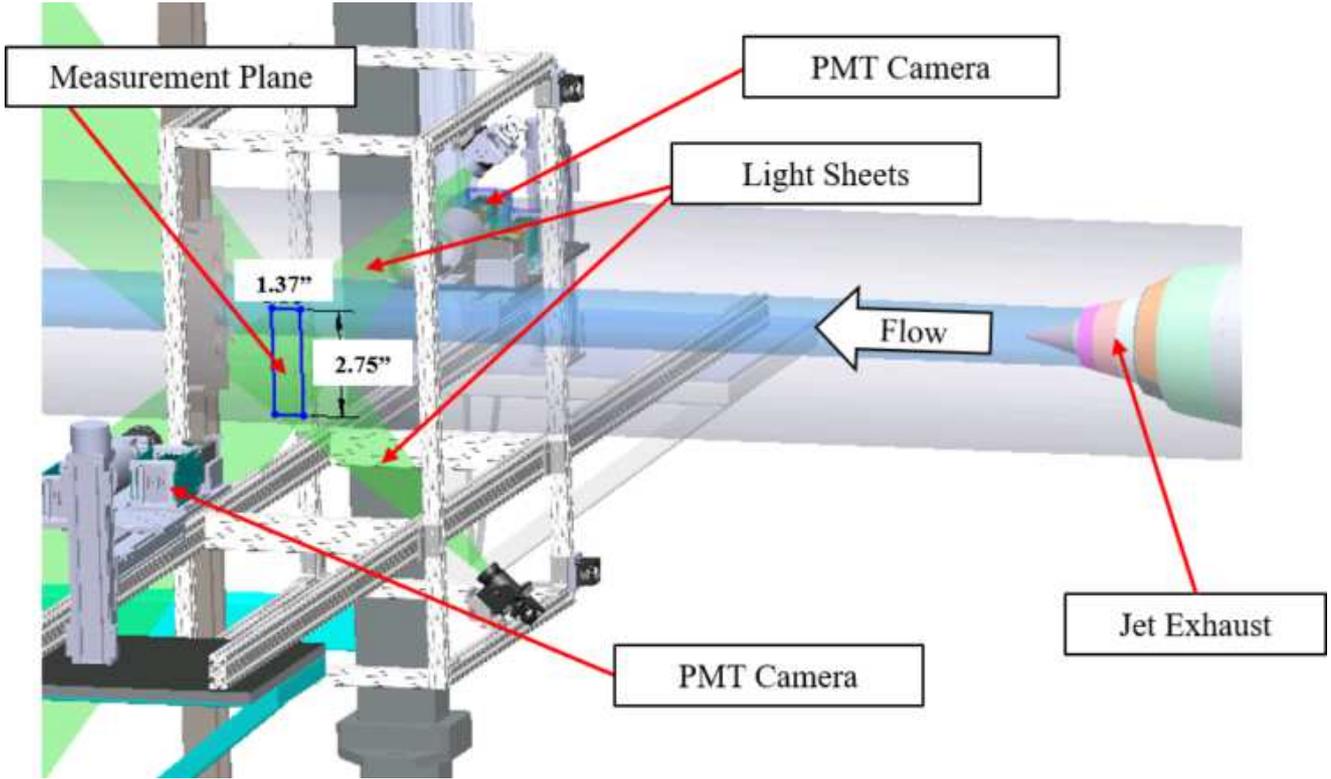}
\caption{DGV instrument installed at AAPL in the NATR.}
\label{fig:1}
\end{figure*}

The DGV system uses two laser planes and two photomultiplier tube (PMT) array detectors. The flow is probed non-intrusively using uses a tunable, continuous wave diode-pumped solid state laser frequency doubled to 532 nm wavelength (Coherent Verdi V18). Two overlapping laser sheets create the planar measurement region, as shown in Figure \ref{fig:1}. The light sheets are multiplexed with a time difference of 2 $\mu s$, and the flow was sampled for 1 s with a sampling rate of 250 kHz. The multiplexing of the laser sheets was accomplished using two IntraAction Corp. 80 MHz acousto-optical modulators. \

 \begin{figure*}
\centering
  \includegraphics[width=0.75\textwidth]{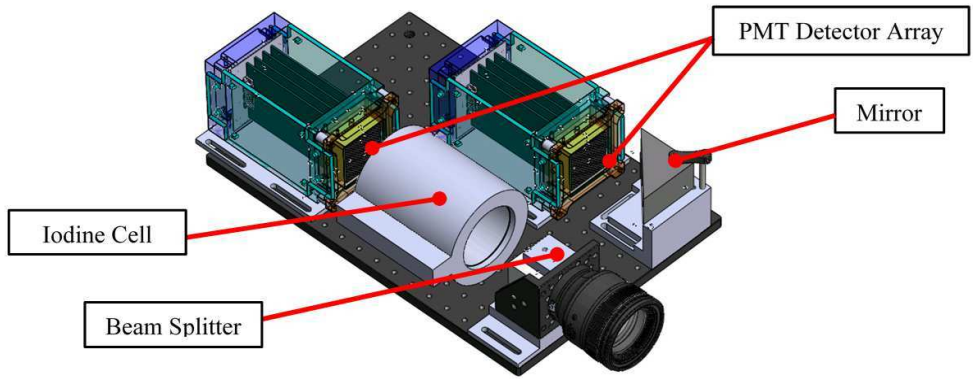}
\caption{PMT camera assembly.}
\label{fig:2}
\end{figure*}

The multichannel PMT cameras consist of two Hamamatsu H8500C PMT array detectors with 64 channels on each detector. For this experiment only 32 channels were used on each detector with 4 channels spaced in the axial direction and 8 channels in the radial direction. The PMT detectors were oriented to collect light scattered perpendicular to the measurement plane, with one on each side of the measurement plane (see Figure \ref{fig:1}). As shown in Figure \ref{fig:2}, the light entering the PMT camera is split with one path entering an iodine cell before reaching the PMT array detector. The other path is redirected with a mirror and goes directly to the second PMT array detector. The focal length of the camera lens used in this experiment was 200 mm, resulting in a magnification of approximately 1.4. This imaging configuration allowed for a measurement plane with dimensions of 34.88 mm (1.37”) in the axial direction and 69.76 mm (2.75”) in the radial direction.\

\begin{figure*}
\centering
\begin{subfigure}[b]{0.3\textwidth}
        \includegraphics[width=\textwidth]{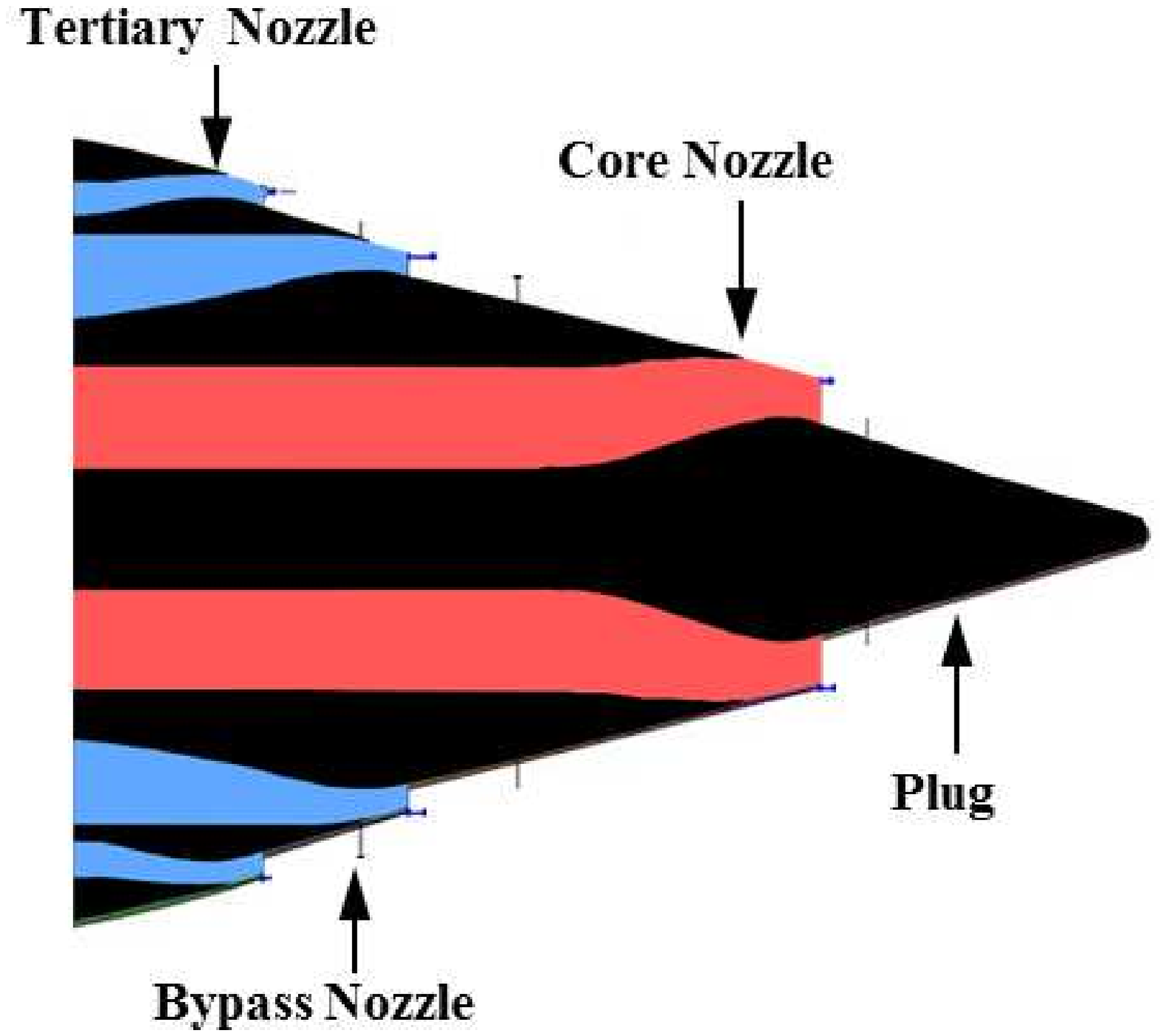}
        \caption{ }
        \label{fig:3a}
    \end{subfigure}
    \begin{subfigure}[b]{0.3\textwidth}
        \includegraphics[width=\textwidth]{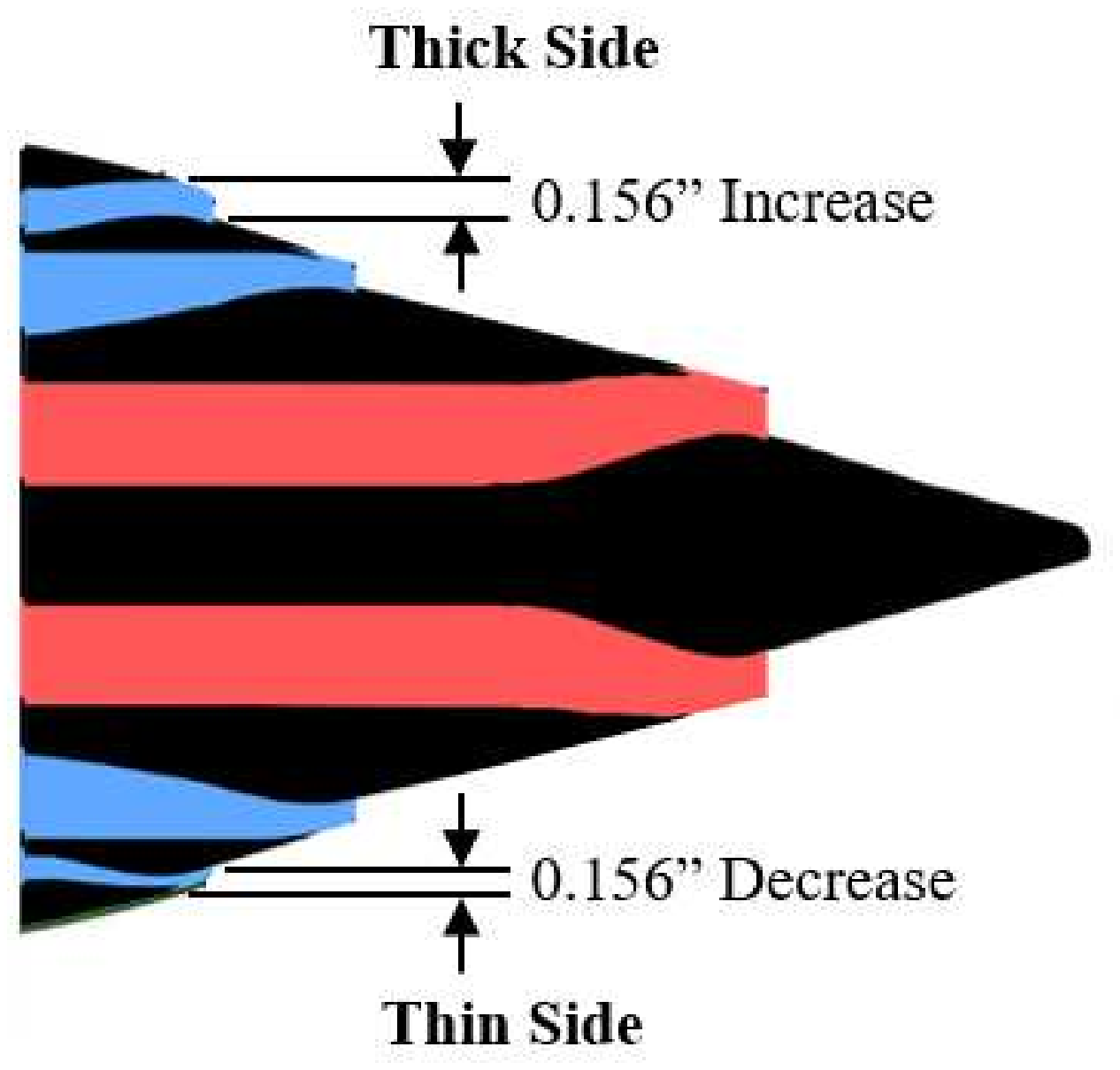}
        \caption{ }
        \label{fig:3b}
    \end{subfigure}
 
\caption{Three-stream nozzle configurations (a) axisymmetric (b) asymmetric with tertiary offset.}
\label{fig:3}
\end{figure*}

The experiment was conducted at AAPL at the NASA Glenn Research Center. The three-stream nozzle configurations were installed on the High Flow Jet Exit Rig (HFJER) in the Nozzle Acoustic Test Rig (NATR). The nozzles used are externally-mixed and externally-plugged as shown in Figure \ref{fig:3}. The red area in the figure denotes the heated, core flow. The blue areas denote the bypass and tertiary flows, which were heated to a much lower total temperature than the core. The area ratio of the bypass to the core was 2.5, and the area ratio of the tertiary to the core was 1.0. The subscripts, c, b, and t are used to denote the core, bypass, and tertiary streams, respectively. These dimensions are the same for both the axisymmetric and asymmetric configurations. Typically the diameter of the nozzle is used to present results in non-dimensional units; however, since the nozzles studied are externally-plugged, an equivalent diameter is calculated based on the nozzle area. The equivalent diameter, $D_{eqA}$, is based on the total exhaust area of the three-stream nozzle.  To create the asymmetry nozzle configuration, the tertiary stream was offset by 4 mm (0.156”), which is approximately 2\% of $D_{eqA}$ (see Figure \ref{fig:3}(b)), resulting in a locally “thick” shear layer on that side [11].\

The flow was seeded using three different systems - one for the core, one for the bypass, and one for the tertiary stream - but each using a pH-stabilized aluminum oxide dispersion (\cite{wernet1994stabilized}). These particles scatter the laser light when in the field of view of the DGV PMT cameras, providing the signal used for further processing. The same seeding system was used for PIV measurements.\

Table \ref{tab:1} lists the operating conditions of the jet for the configurations tested. The nozzle pressure ratio (NPR) is defined as the total pressure divided by the ambient pressure. The nozzle temperature ratio (NTR) is the total temperature divided by the ambient temperature. For the axisymmetric nozzle configuration, PIV data were provided by \cite{henderson2016characterization} to allow comparisons and synthesis of these mean flow and turbulence results with convection velocity results from the Virginia Tech DGV instrument. \

\begin{table*}
\centering
\caption{Operating conditions.}
\label{tab:1}       
\begin{tabular}{cccccccc}
\hline\noalign{\smallskip}
  & $NPR_c$ & $NPR_b$ & $NPR_t$ & $NTR_c$ & $NTR_b$ & $NTR_t$ & Instrument\\
\noalign{\smallskip}\hline\noalign{\smallskip}
Axisymmetric&	1.8	&1.8&	1.4&	3.0&	1.25&	1.25&	DGV/PIV\\
Offset&	1.8	&1.8&	1.4&	3.0&	1.25&	1.25&	DGV\\

\noalign{\smallskip}\hline
\end{tabular}
\end{table*}

\begin{table*}
\centering
\caption{Jet conditions.}
\label{tab:2}       
\begin{tabular}{ccccc}
\hline\noalign{\smallskip}
$U_j (m/s)$&	$a_j   (m/s)$&	$a_\infty (m/s)$	&$M_j$&$	Re$\\
\noalign{\smallskip}\hline\noalign{\smallskip}
519&	542	&335&	0.956&	$1.4 \times 10^6$\\

\noalign{\smallskip}\hline
\end{tabular}
\end{table*}

Table 2 lists the jet conditions. The subscript $j$ denotes conditions at the exit of the core stream while the subscript $\infty$ denotes the ambient conditions. The Reynolds number is based on the conditions at the exit of the core stream and uses $D_{eqA}$ as the characteristic length. The conditions are the same for both the axisymmetric and the offset nozzle configurations.\

The DGV instrument was mounted to a large traverse to sample the flow at many axial and radial locations to measure the majority of the developing jet plume. For all tested conditions, the instrument was moved to 4 distinct radial locations. As shown in Figure \ref{fig:1} the measurement plane was 34.88 mm (1.37 inches) in the axial direction by 69.76 mm (2.75 inches) in the radial direction for the 32 pixel sensor.\

\section*{Analysis}
A new method utilizing the measured laser scattering from particles seeding the flowfield was used to calculate convection velocity of the particle concentration field. This new method differs from that used previously by \cite{ecker2015eddy} who instead used the Doppler-shifted velocity signal. The particle scattering signal is proportional to the instantaneous concentration of particles in the flowfield, and in the limit of very short time scales will exactly follow the velocity field (i.e., the principal assumption of PIV). At longer timescales, this signal will contain additional nonlinear terms dominated by vortical transport of the particle field. This technique relies upon one of many physical quantities used previously for obtaining a measure of the convection speed of large-eddy features in turbulent flows.\
 \begin{figure}
\centering
  \includegraphics[width=0.4\textwidth]{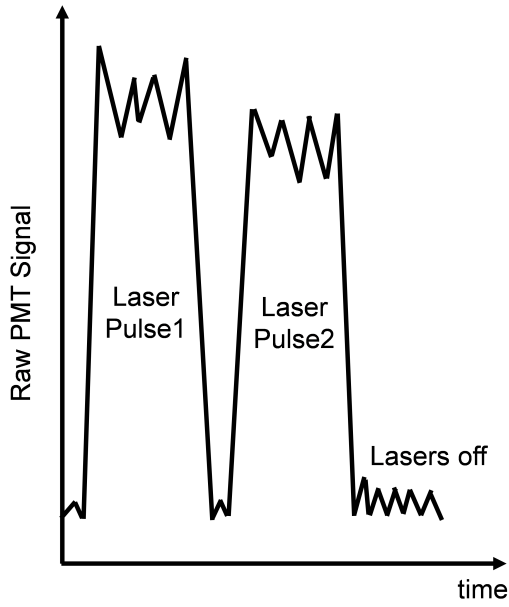}
\caption{Example of raw signal from unfiltered PMT detector.}
\label{fig:4}
\end{figure}
The particle signal is extracted from the raw signal recorded by the PMT detector array without the iodine cell (see Figure \ref{fig:2}, the array depicted on the right). The scalar signal contains information about the turbulent structures because the bursts of seed particles are convected with the turbulence in the jet flow. Due to the multiplexing of the laser sheets, the raw signal contains data when both lasers were on and off. An example of a pair of laser pulses from the raw signal is shown in Figure \ref{fig:4}. For each raw signal from the unfiltered PMT detector, two scalar signals can be obtained, one for each laser. The scalar signal for each laser was extracted from the PMT detector raw signal by taking the mean value during the laser pulse emissionand subtracting from it the mean value when the laser was off. This was done for each laser pulse to obtain a scalar signal for each laser with a record length of 250,000 samples. \

Through the use of the DGV system, the scalar signal was obtained for each of the 32 pixels simultaneously at each axial location. A second-order space/time correlation function can be calculated using the following definition (\cite{morris2010velocity}): \

\begin{equation}
\begin{split}
R_{ij}(x,y,t,\xi,\eta,\tau) \\
=\frac{1}{T} \int_{-T/2}^{T/2} s_i(x,y,t)s_j(x+\xi,y+\eta,t+\tau)dt \\
=\overline{s_i(x,y,t)s_j(x+\xi,y+\eta,t+\tau)}
\end{split}
\end{equation}

where $T$ is the data acquisition time period, $s$ is a generic flow fluctuation variable (in this case the fluctuation of a passive scalar), $\xi$ is the correlation streamwise spacing, $\eta$ is the correlation radial spacing, $\tau$ is the correlation lag time, and the bar denotes time averaging. 
The convection velocity of the eddies can be calculated using the information from the second-order space/time correlation function, as shown by  \cite{ecker2015eddy}. The correlation is calculated at each $x/D$ location, for each $\Delta x/D$ spacing provided by the four streamwise pixels at a given $r/D$ location in the flow. Therefore, for each $x/D$ location, the space/time correlations across the sensor in the axial direction can all be plotted on one figure with $\tau$ as the independent variable. \

\begin{figure*}
\centering
  \includegraphics[width=0.75\textwidth]{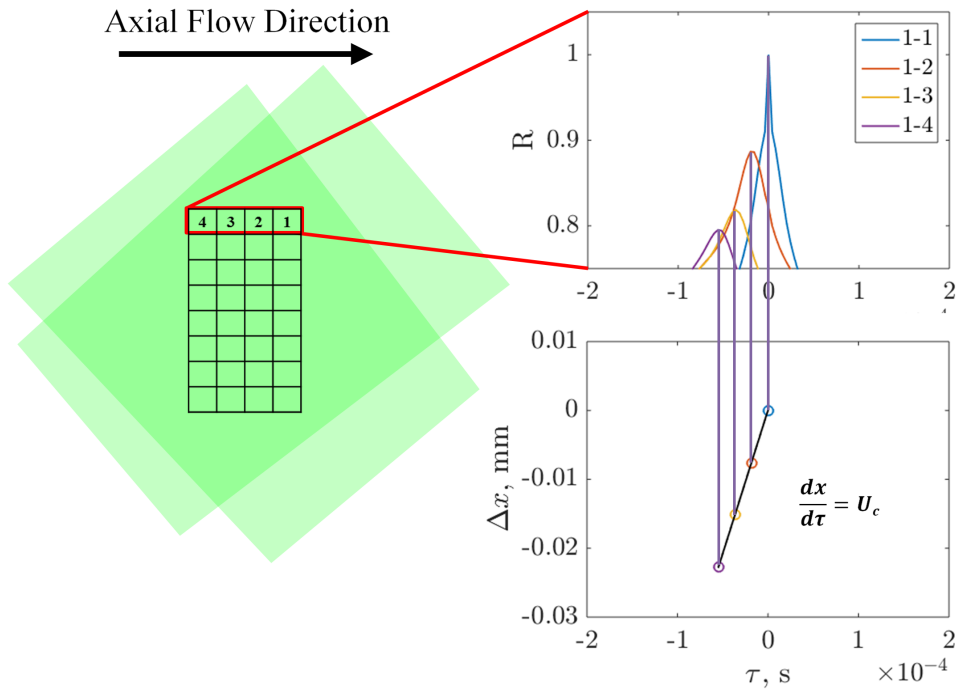}
\caption{Schematic of processing for calculating convection velocity.}
\label{fig:5}
\end{figure*}

Since the correlation function has maxima along the convection ridge, the equation $\xi_r = U_c \tau_r$ can be used to solve for the convection velocity, $U_c$. The spacing between pixels, $\Delta x$, is known (\cite{stuber2017investigation}); so, the equation can be rewritten as $U_c=N \Delta x/\tau_r$. The slope, $\Delta x/\tau_r$, can be determined from the aforementioned plots of the correlation functions (\cite{fisher1964correlation}). Extracting the maxima of the correlation curves using a cubic spline interpolation and plotting the correlation lag time on the $x$-axis versus spacing on the y-axis allows a linear least-squares fit to be performed on the four points. The resulting slope of this line is the measured convection velocity. This process is shown in Figure \ref{fig:5}, where the correlation functions depicted are examples taken from the current data. As can be seen in Figure \ref{fig:5}, the lag times versus spacing is linear, which supports our convention velocity measurement method.  The uncertainty in the calculation of the convection velocity is $\pm0.07 \delta U_c/U_c$. Further details on the uncertainty analysis are described by \cite{stuber2017investigation}. \

The above explained method of calculating the convection velocity from the space/time cross-correlation of the scalar signal is different than common velocity measurement techniques, such as PIV, which measures the true velocity of the flow. Although both the proposed method for calculating the convection velocity and PIV use space/time correlations of particle scattering-based signals, they provide measurements of different velocities of the flow. In PIV the time delay over which space/time correlations are measured is small compared to the time scales of interest in the flow. Because of the small time delay, Taylor’s hypothesis is valid within a very small truncation error in time, resulting in equivalence between convection velocity and the true velocity. For the particle signal from the DGV instrument, the space/time correlations are over a much longer time (milliseconds versus microseconds), and Taylor’s hypothesis is no longer generally valid. Since the space/time cross correlations are over a larger time, the velocity calculated includes nonlinear effects of convection, allowing the convection velocity on the convective ridge to be calculated.\

\section*{Results and Discussion}
In the discussions to follow, the convection velocity data obtained from the DGV technique are interpreted for three-stream nozzle configurations—one axisymmetric and one with an offset third stream intended for noise reduction. Prior turbulence intensity results obtained by \cite{henderson2016characterization} are considered along with observations on the mean convection velocity structure to postulate some key mechanisms of significance to noise radiation in these configurations. \

\label{sec:1}
\subsection*{Axisymmetric Nozzle Configuration}
\begin{figure*}
\centering
  \includegraphics[width=1.00\textwidth]{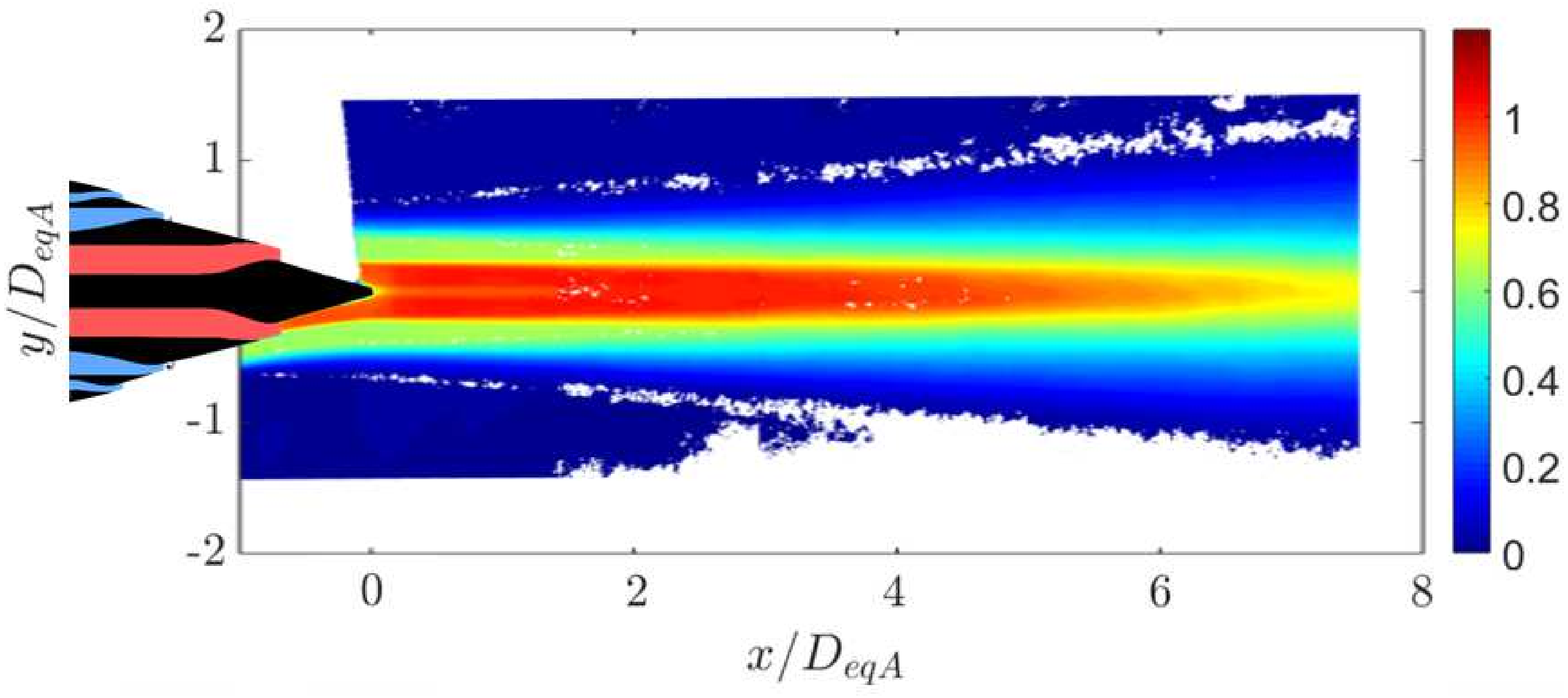}
\caption{NASA PIV mean velocity, $U/U_j$, for the axisymmetric nozzle configuration.}
\label{fig:6}
\end{figure*}

\begin{figure*}
\centering
  \includegraphics[width=1.00\textwidth]{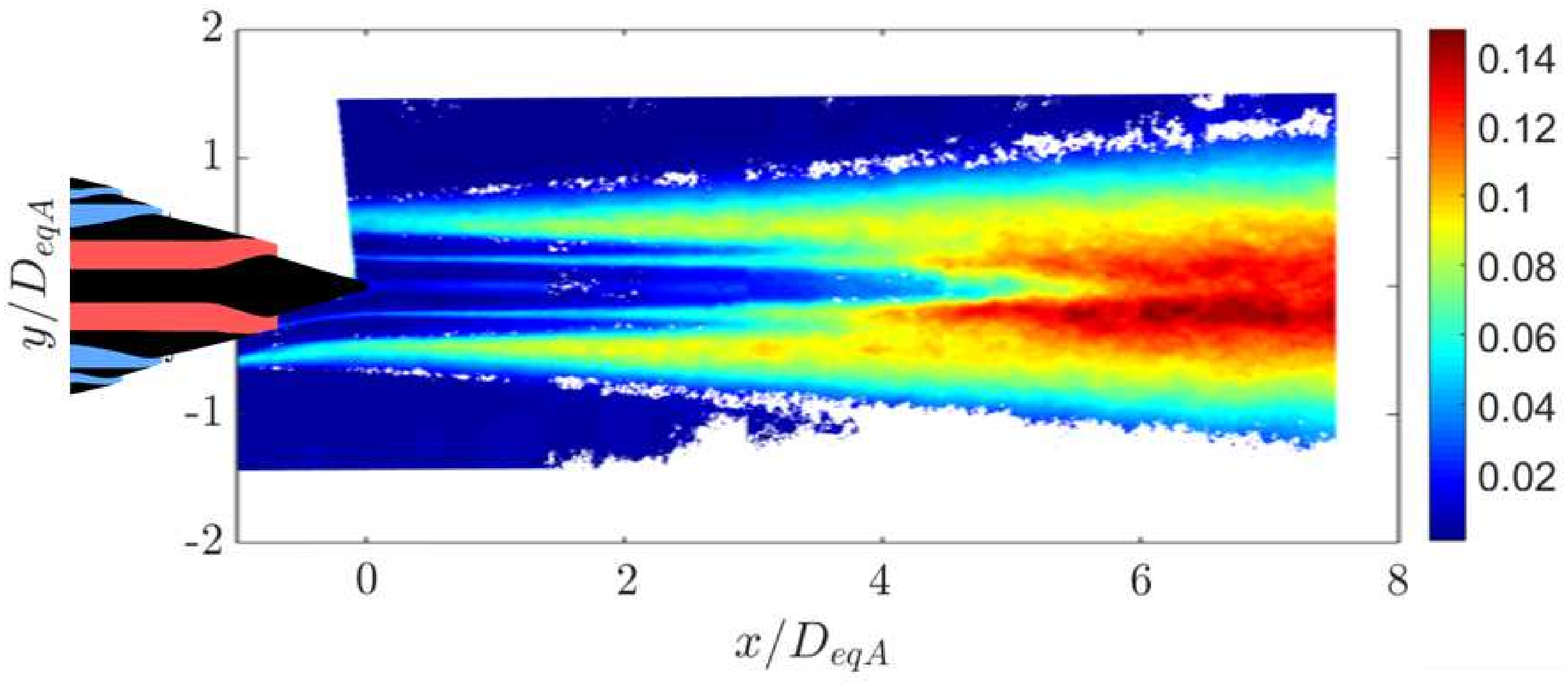}
\caption{NASA PIV turbulence intensity, $u^\prime /U_j$, for the axisymmetric nozzle configuration.}
\label{fig:7}
\end{figure*}

For basic insight on the baseline flow field for the axisymmetric configuration, the NASA PIV mean velocity and turbulence intensity distributions (\cite{henderson2016characterization}) are shown in Figures \ref{fig:6} and \ref{fig:7}, respectively. Both the mean velocity and turbulence intensity data are nondimensionalized by the isentropic core exit velocity, $U_j$. The axial and radial coordinates have been nondimensionalized by the equivalent diameter based on the total exit area of the nozzle, $D_{eqA}$. As the flow exits the nozzle, the core and the bypass streams are visible in the mean velocity field (Figure \ref{fig:6}). The flow exits the nozzle in the core at a velocity very close to the isentropic exit velocity and maintains this velocity until about 5 area-equivalent diameters downstream.\

The potential core regions of the core and bypass streams are clearly visible in Figure \ref{fig:7}, as the potential core is characterized by the low turbulence intensity region. From the PIV data it is observed that the tertiary stream seems to mix with the bypass stream at an early axial location, as a distinct tertiary stream is not visible in either the turbulence intensity or the mean velocity distributions. Based on the turbulence intensity data, the mean potential core is overcome by the inward-growing shear layer just after $5D_{eqA}$ downstream, which is where the turbulence intensity along the centerline begins to increase.\

\begin{figure*}
\centering
  \includegraphics[width=1.00\textwidth]{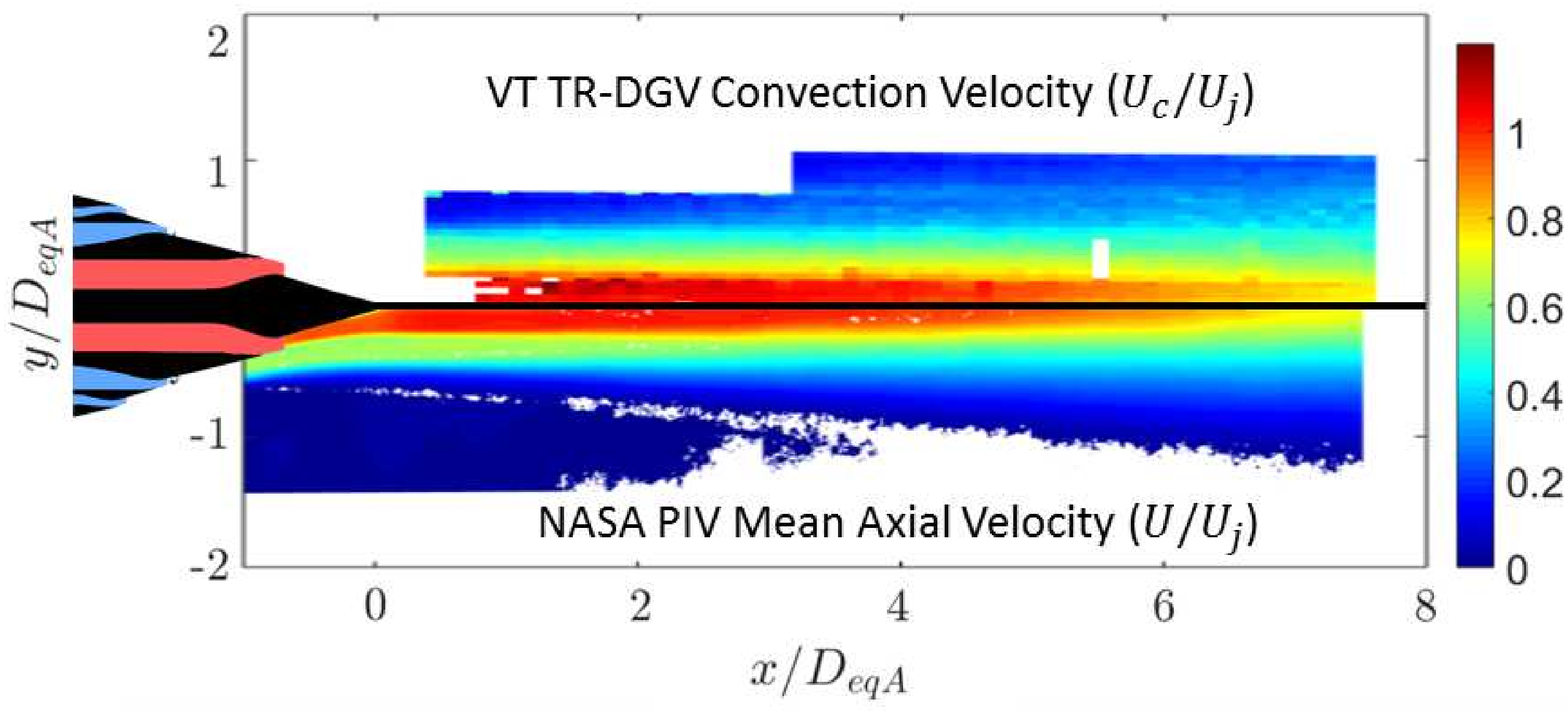}
\caption{Comparison of VT DGV convection velocity (values plotted for $y/D_{eqA} >0$) to NASA PIV mean velocity (values plotted for $y/D_{eqA} <0$), for the axisymmetric nozzle configuration.}
\label{fig:8}
\end{figure*}

The VT DGV convection velocity data are compared to the NASA PIV mean velocity in Figure \ref{fig:8}. The convection velocity data have been nondimensionalized by the isentropic exit velocity. Again, the axial and radial coordinates have been nondimensionalized by $D_{eqA}$. An axial lag between the VT DGV convection velocity and the NASA PIV mean velocity is observed in Figure \ref{fig:8}, particularly near the end of the potential core. Even though the mean velocity decreases along the centerline of the jet, the convection speed of the turbulent structures persists for a longer distance. Although more noticeable at larger axial locations, the convection velocity tends to be greater than the mean velocity in the outer portion of the shear layer. This trend has been observed in previous studies (\cite{williams1965mach}; \cite{papamoschou2010beamformed}; \cite{petitjean2007space}) and is attributed to the entrainment of turbulent eddies with the shear layer which re-energize the shear layer.\

\begin{figure*}
\centering
\begin{subfigure}[b]{0.45\textwidth}
        \includegraphics[width=\textwidth]{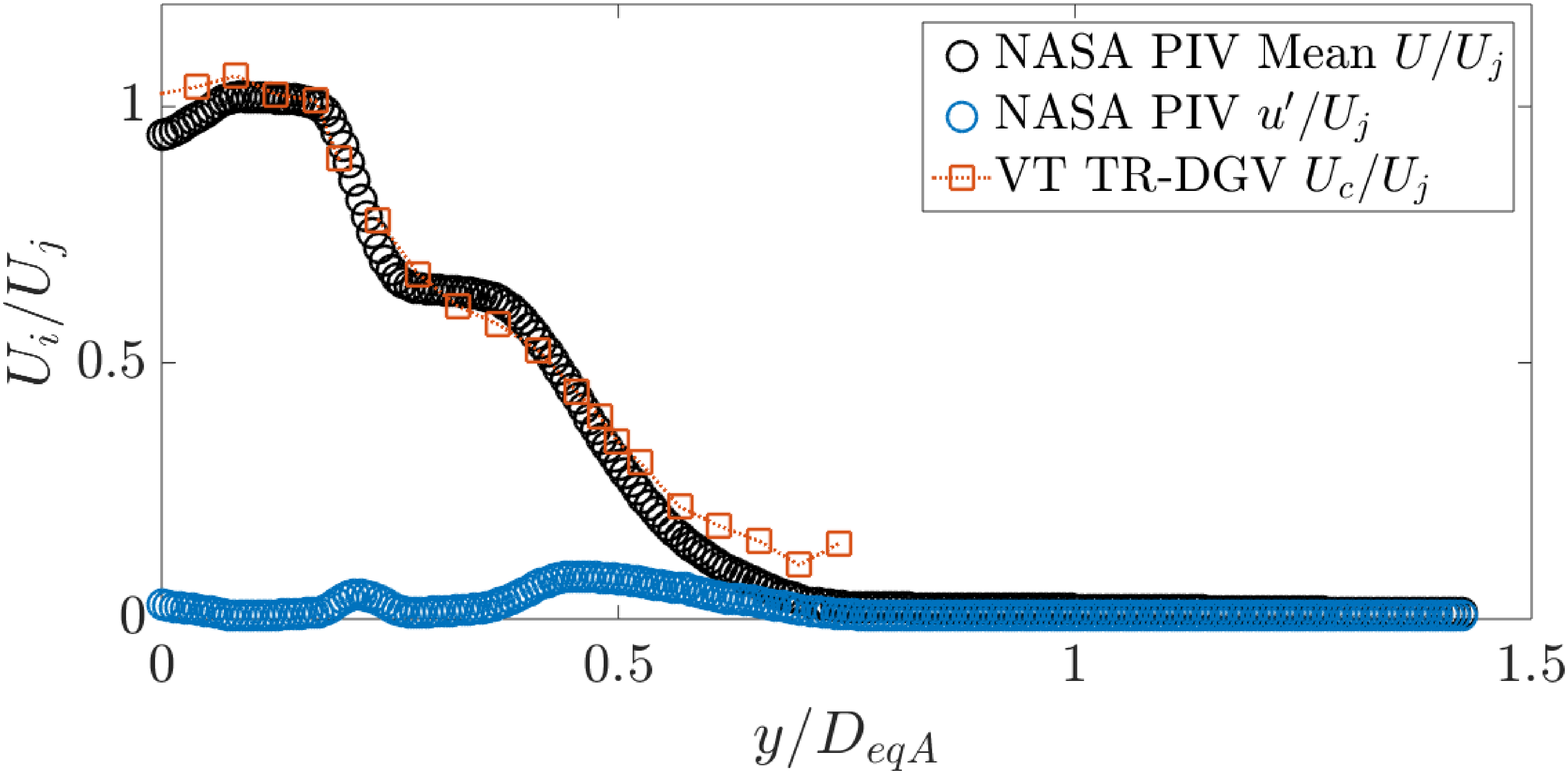}
        \caption{ }
        \label{fig:9a}
    \end{subfigure}
    \begin{subfigure}[b]{0.45\textwidth}
        \includegraphics[width=\textwidth]{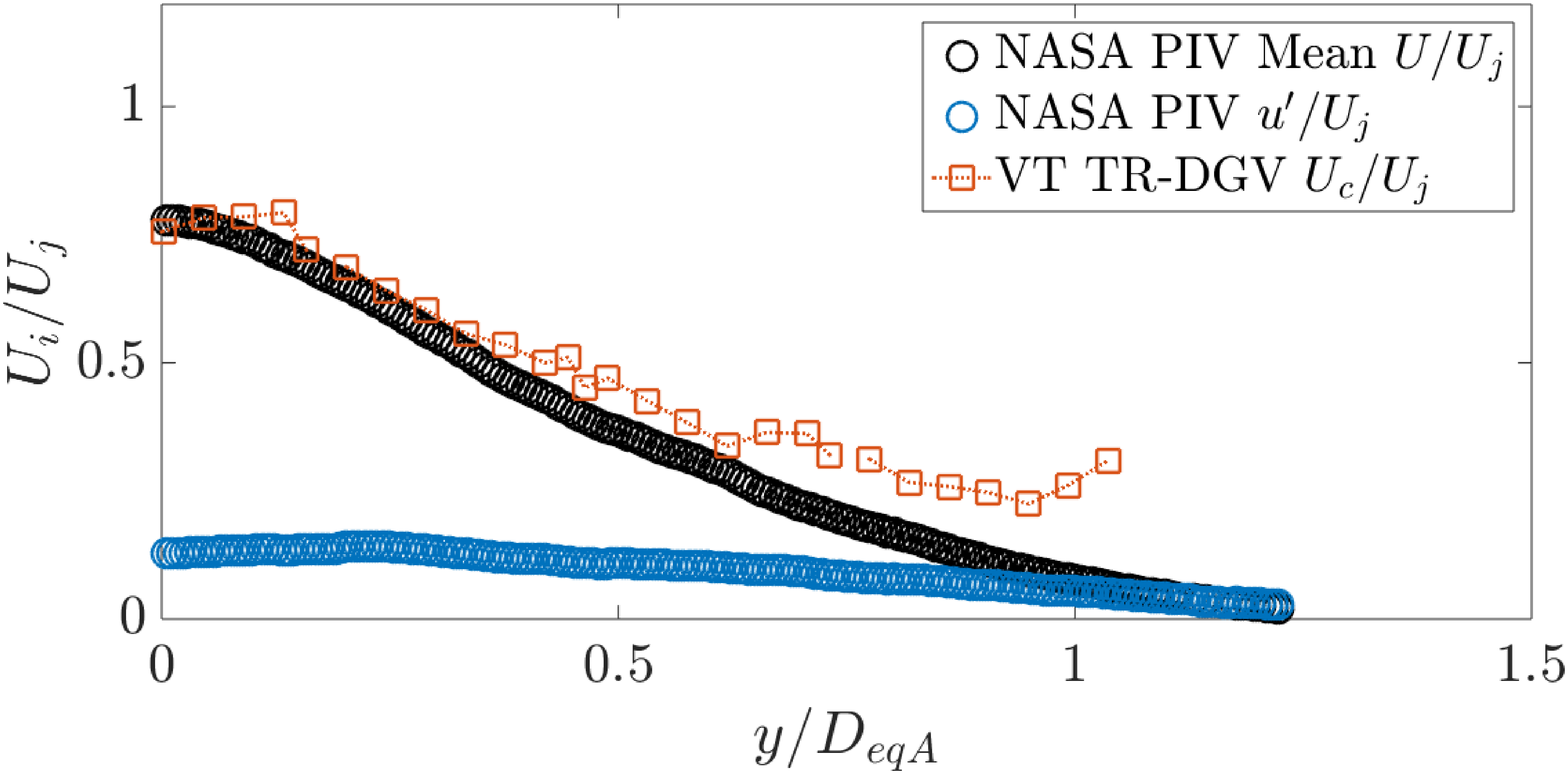}
        \caption{ }
        \label{fig:9b}
    \end{subfigure}
 
\caption{Radial contours at (a) $x/D_{eqA}=1.0$ and (b) $x/D_{eqA}=7.0$.}
\label{fig:9}
\end{figure*}

Radial profiles of the NASA PIV Mean Velocity, NASA PIV turbulence intensity and VT DGV convection velocity are shown in Figure \ref{fig:9} for more quantitative comparisons of the shear layer features. Near the nozzle exit and close to the centerline, the convection velocity follows the mean velocity, as previously observed by \cite{bridges2017measurements}. At greater radial locations, the convection velocity deviates from the mean velocity, as already mentioned with respect to Figure \ref{fig:8}. Statistically, the deviation is due to higher-velocity flow closer to the core axis being dominant in the determining the peak correlation function values in these regions. Physically, this result reflects the influence of the core-flow-generated turbulence in setting the local wavespeeds near the edges of the shear layer.\

In contrast to the near-nozzle region, the radial profiles at $D_{eqA} =7$, downstream of axial stations with a mean potential core region, indicate that the local convection velocity is measurably greater than the local mean velocity. These results are consistent with those reported by \cite{ecker2015eddy} and \cite{shea2017eddy}. \cite{bridges2017measurements} analyzed convection velocities measured with time-resolved PIV and considered results from the literature for near-field hydrodynamic pressure convection velocity to interpret the findings. They concluded that local convection velocity was equivalent to the local mean fluid velocity, and that the resultant nearfield hydrodynamic pressure convection velocity was simply the mean velocity in the shear layer at the location of peak turbulence intensity. \cite{papamoschou2018modelling} has also considered the relationships between nearfield hydrodynamic pressure and flowfield features, using large-eddy simulation results to determine that the peak magnitude of Reynolds shear stress better captures the position with the same mean streamwise velocity as the convection velocity. The current results are largely consistent with the conclusions of \cite{bridges2017measurements} regarding similarity between local mean fluid velocity and local mean eddy convection velocity, with two exceptions. First, as already discussed, the flow on the outer edges of the shear layer differ due to the dominant fluctuation energies of eddies fed by the higher velocity flow nearer the jet axis. Second, the convection velocity immediately downstream of the end of a potential core annulus is consistently greater than the local mean fluid velocity there. This second difference is thought to be significant, given the large turbulence intensities in these regions. Since offset streams modify the mean flow and turbulence structure around the potential core breakdown region, the structure of the convection velocity is of particular interest for comparison across axisymmetric and offset cases.\

As shown through Lighthill’s acoustic analogy and the Reynolds stress tensor, areas of large velocity fluctuations are important for noise generation (\cite{lighthill1952sound}). In order for aft-radiated, large-scale disturbances in the flow to reach the far field, however, they must occur at wavenumbers and frequencies which have supersonic phase speeds with respect to the ambient medium (\cite{jordan2013wave}). Therefore, not only are areas of high turbulence intensity of interest, but also areas of high convection speeds are of interest for noise generation in jet plumes. Further, the radiation efficiency relationship developed by \cite{papamoschou2014reduction} showed that both the turbulence intensity and convection velocity contribute the amount of radiated noise in jets. Therefore, comparison of the convection velocity distribution to the turbulence intensity distribution offers insight into regions of the jet most likely to be important to noise radiation. \
 	
\begin{figure*}
\centering
  \includegraphics[width=0.75\textwidth]{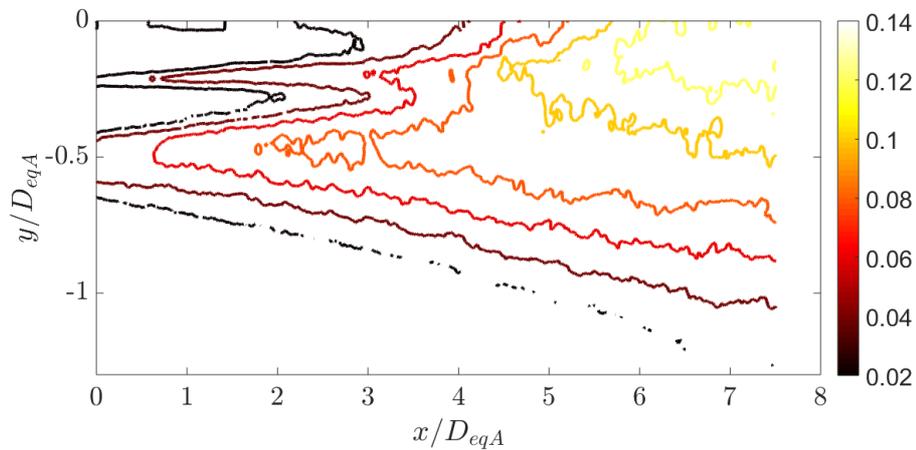}
\caption{Contours of the NASA PIV turbulence intensity measurements, $u^\prime/U_j$, for the axisymmetric case.}
\label{fig:10}
\end{figure*}

\begin{figure*}
\centering
  \includegraphics[width=0.75\textwidth]{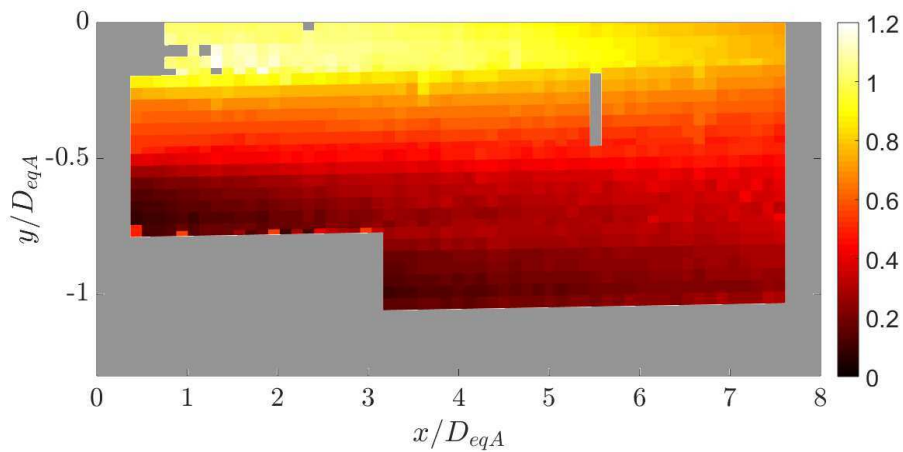}
\caption{Distribution of VT DGV convection velocity, $U_c/U_j$, for the axisymmetric case.}
\label{fig:11}
\end{figure*}

\begin{figure*}
\centering
  \includegraphics[width=0.75\textwidth]{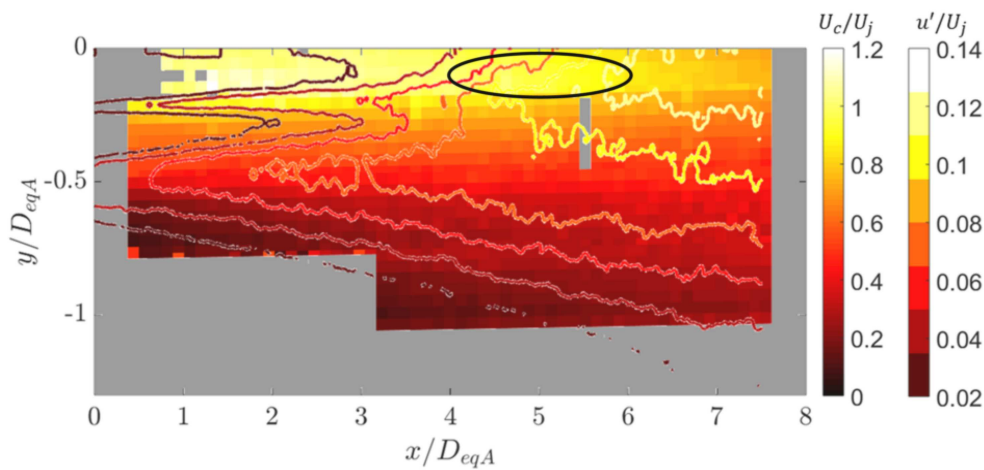}
\caption{Turbulence intensity (lines) and convection velocity (fill) for axisymmetric case. The black oval indicate the region of interest for intense noise production.}
\label{fig:12}
\end{figure*}

To make this comparison, consider the overlay of the NASA PIV turbulence intensity distribution contours (Figure \ref{fig:10}) on the VT DGV convection velocity data (Figure \ref{fig:11}) in Figure \ref{fig:12}. Note that the background of Figure \ref{fig:11} is colored gray where no data are available. From the overlaid contours, areas of both high convection velocity and high turbulence intensity are seen downstream of the end of the potential core, from $4\leq x/D_{eqA} \geq 6$ and offset from the centerline around $y/D_{eqA}= -0.1$ (black oval in Figure 12). These findings suggest this region along the centerline is an important location in the jet plume for noise production. Further, the identified region is consistent with flowfield/farfield correlation measurements by \cite{panda2005investigation}, who identified the same region downstream of the end of the potential core as the strongest sound producing source for high subsonic and supersonic single-stream jets.\

 \label{sec:2}
\subsection*{B.	Offset Nozzle Configuration}
\begin{figure*}
\centering
  \includegraphics[width=0.75\textwidth]{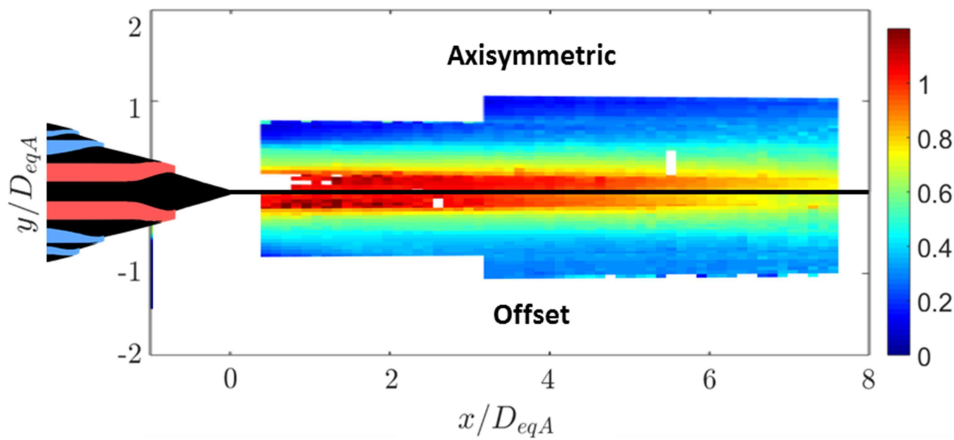}
\caption{Comparison of convection velocity, $U_c/U_j$, (top) axisymmetric nozzle configuration and (bottom) offset nozzle configuration (“thicker side”).}
\label{fig:13}
\end{figure*}

In the following section only convection velocity results from the VT DGV instrument are presented in order to indicate the differences in convection velocity structure caused by the offset third-stream geometry. For the offset nozzle configuration, all results are for the locally “thicker” side of the jet; i.e., on the side with the wider third-stream nozzle exit (see Figure \ref{fig:3}(b) at the top). Comparison of the three-stream axisymmetric nozzle convection velocity with that of the three-stream asymmetric offset case shown in Figure \ref{fig:13} reveals clear differences between the flow fields. The convection speeds at large radial locations (i.e., outer layer of the jet) are higher for the offset case. The increase in momentum due to the thicker shear layer is most likely responsible for this increase in convection velocity, which promotes more rapid transfer of momentum from the jet to the ambient. Along the centerline, however, convection speeds for the offset configuration begin to decrease at an earlier axial location. This is evidence that offset causes the potential core to breakdown earlier compared to the axisymmetric configuration, and this breakdown impacts the convection velocity to a measurable significance. \
 
\begin{figure*}
\centering
  \includegraphics[width=0.75\textwidth]{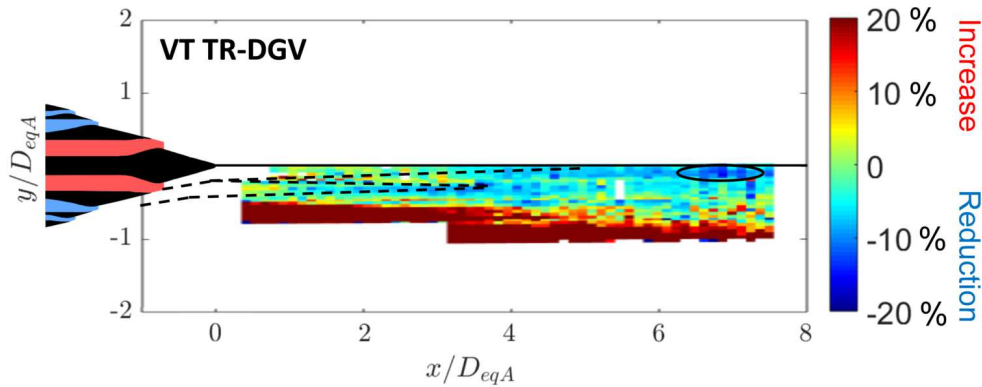}
\caption{Axisymmetric/offset convection velocity percent difference distribution. Dashed lines: potential cores; Oval: largest reductions.}
\label{fig:14}
\end{figure*}

Considering these differences further, the percent reduction/increase in the local convection velocities between the offset and axisymmetric cases are shown in Figure \ref{fig:14}. Positive values indicate an increase in convection velocities, while negative values indicate a reduction in convection velocities due to the offset configuration. The dotted black lines indicate the location of the potential core boundaries for the core and bypass streams. Two regions are readily identified as reduction areas: one in the shear layer and another along the centerline. The reduction in the shear layer region is expected as the introduction of a greater momentum stream will enhance the mixing and, therefore, entrainment of the ambient fluid to slow convection speeds (indeed, this is reflected inversely in the increases seen at the radial edges of the measurement region). The region from $4\leq x/D_{eqA} \geq 6$  and slightly offset from the centerline around $y/D_{eqA} = -0.1$ shows the greatest reduction in convection velocities and is also in the vicinity of the end of the potential core. Specifically, the area around $x/D_{eqA}=7$ shows a reduction by almost 20\%.\

Computational work by \cite{papamoschou2012modeling} and experimental work by \cite{papamoschou2017very} have shown that noise reduction in asymmetric jets is caused in part by a reduction in the convection Mach number. Recent work (\cite{mayo2017experimental}; \cite{mayo2018flow}); \cite{daniel2018experimental}) has also shown that temperature non-uniformities can produce reductions in sound pressure levels due to a mean flow and turbulence structure modification. Work by \cite{veltin2011correlation} showed two areas of high correlation for jet noise, one before and one after the end of the potential core. It is likely that the two areas of reduction may be the two noise sources identified by Veltin et al. It is plausible that the introduction of the offset reduces the noise generation in part through a reduction in the largest convection velocities both in the shear layer and along the centerline. \

As already noted, some regions in the flow have increased convection velocity, but these exhibit convection velocities of at-most $0.2U_j$ and would not impact the sound radiated to the far field. Rather, as recent computational work by \cite{unnikrishnan2017acoustic} suggests, these effects may be important for energy dissipation without radiation. They show that vortex intrusion from the shear layer into the core is responsible for noise generation, while entrainment of ambient and shear layer fluid into the core causes shear-driven energy dissipation. The net interaction of these two mechanisms shows dissipation within the potential core while production in the shear layer and beyond the end of the potential core. Increased entrainment of the ambient fluid is observed in the offset configuration due to the higher convection speeds in the outer regions of the shear layer. A possible physical explanation of the noise reduction due to the offset, therefore, is that increased entrainment of the ambient air causes a stronger dissipation within the core, resulting in an earlier collapse of the potential core. The earlier collapse reduces the convection velocity along the centerline earlier and reduces the extent and intensity of the noise production region beyond the end of the potential core and in the shear layer.\

Additionally, the area along the centerline identified as having the greatest reduction in convection velocity for the offset case is the same area identified as having the highest convection velocities and turbulence intensity for the axisymmetric case in Figure \ref{fig:12}. The reduction in convection velocity in this region may be partially responsible for the reduction in radiated noise on the thick side measured by \cite{henderson2015measurements}. It is also evident that the tertiary stream increases the convection velocity in the outer portion of the jet.\

%

\bigskip
\section*{Conclusions}
\label{conc}
Flow features linked to noise radiation were investigated for heated, three-stream nozzle configurations. Both an axisymmetric and an offset nozzle configuration were tested. The TTR for the core was 3.0 and for the bypass and tertiary streams was 1.25. The NPR for the core and bypass streams was 1.8, and the NPR for the tertiary stream was 1.4. Mean velocity and turbulence intensity from PIV measurements made by researchers at NASA Glenn Research Center were analyzed in light of convection velocity measured using the Virginia Tech time-resolved Doppler global velocimetry instrument (VT DGV). Multichannel detectors collected scattered laser light from particles following the jet flow at 250 kHz. Simultaneous measurements from 32 planar points in the flow allowed for calculation of the mean convection velocity of the particle concentration field using a statistical cross-correlation on the scalar signal along the four pixels in the axial direction. The convection velocity is the slope of the line connecting the times of maximum correlation across axial spacing.\

For the axisymmetric nozzle configuration, comparison of the VT DGV convection velocity to the NASA PIV mean velocity measurements show the convection velocity lags the mean velocity along the centerline. Comparisons of the VT DGV convection velocity with turbulence intensity show the region along the centerline, from 4 to 6 nozzle diameters downstream, have both high turbulence intensity and convection velocity. This finding is an indication that this region downstream of the end of the potential core may be an important region for noise generation and is consistent with a number of past conclusions about noise source origins. Further, comparison of the axisymmetric and offset convection velocity reveal that, in addition to the shear layer region, the same region along the centerline is the area of greatest reduction in convection velocity. These reductions in convection velocity offer a compelling explanation to reduction in noise on the “thick” side of the offset third stream observed in acoustic measurements. \

The observations made herein, while informed by a number of studies in the literature, should be bolstered in their explanation though further work. Specifically, near-field pressure measurements could be used to immediately measure the importance of the post-potential core convection speeds on radiated noise in these flows. Further, additional statistical analysis of the large-scale structural features in the PIV measurements could reveal more insights into the turbulence and instability characteristics of the structures interacting at the end of the potential core. Given these efforts, a clearer picture of the precise mechanisms for offset third stream noise reduction could be obtained, allowing for better exploitation of the concepts for practical noise reduction.\ 

\bigskip

\begin{acknowledgements}
The authors gratefully acknowledge the hard work in collecting the DGV data by Dr. Tobias Ecker. Additionally, the authors gratefully acknowledge Drs. Brenda Henderson and Mark Wernet of the NASA Glenn Research Center for their support and for providing the PIV results. This work is sponsored by Navy grants N00014-16-1-2444 and N00014-14-1-2836, which is funded by the Office of Naval Research and managed by Dr. Steven Martens.\
\end{acknowledgements}
\bigskip

\bibliographystyle{spbasic}      
\bibliography{bibliography_Stuber}{}   

\begin{thebibliography}{31}
\providecommand{\natexlab}[1]{#1}
\providecommand{\url}[1]{{#1}}
\providecommand{\urlprefix}{URL }
\expandafter\ifx\csname urlstyle\endcsname\relax
  \providecommand{\doi}[1]{DOI~\discretionary{}{}{}#1}\else
  \providecommand{\doi}{DOI~\discretionary{}{}{}\begingroup
  \urlstyle{rm}\Url}\fi
\providecommand{\eprint}[2][]{\url{#2}}

\bibitem[{Bridges and Wernet(2017)}]{bridges2017measurements}
Bridges JE, Wernet MP (2017) Measurements of turbulent convection speeds in
  multistream jets using time-resolved piv. In: 23rd AIAA/CEAS Aeroacoustics
  Conference, p 4041

\bibitem[{Daniel et~al.(2018)Daniel, Mayo, Lowe, and
  Ng}]{daniel2018experimental}
Daniel K, Mayo DE, Lowe T, Ng W (2018) Experimental investigation of the very
  near pressure field of a heated supersonic jet with a total temperature
  non-uniformity. In: 2018 AIAA/CEAS Aeroacoustics Conference, p 3145

\bibitem[{Ecker et~al.(2014)Ecker, Brooks, Lowe, and Ng}]{ecker2014development}
Ecker T, Brooks DR, Lowe KT, Ng WF (2014) Development and application of a
  point doppler velocimeter featuring two-beam multiplexing for time-resolved
  measurements of high-speed flow. Experiments in fluids 55(9):1819

\bibitem[{Ecker et~al.(2015)Ecker, Lowe, and Ng}]{ecker2015eddy}
Ecker T, Lowe KT, Ng WF (2015) Eddy convection in developing heated supersonic
  jets. AIAA Journal 53(11):3305--3315

\bibitem[{Ecker et~al.(2016)Ecker, Lowe, and Ng}]{ecker2016scale}
Ecker T, Lowe KT, Ng W (2016) Scale-up of the time-resolved doppler global
  velocimetry technique. In: 54th AIAA Aerospace Sciences Meeting, p 0029

\bibitem[{Fisher and Davies(1964)}]{fisher1964correlation}
Fisher M, Davies P (1964) Correlation measurements in a non-frozen pattern of
  turbulence. Journal of fluid Mechanics 18(1):97--116

\bibitem[{Henderson(2012)}]{henderson2012aeroacoustics}
Henderson B (2012) Aeroacoustics of three-stream jets. In: 18th AIAA/CEAS
  Aeroacoustics Conference (33rd AIAA Aeroacoustics Conference), p 2159

\bibitem[{Henderson and Leib(2015)}]{henderson2015measurements}
Henderson BS, Leib SJ (2015) Measurements and predictions of the noise from
  three-stream jets. In: 21st AIAA/CEAS Aeroacoustics Conference, p 3120

\bibitem[{Henderson and Wernet(2016)}]{henderson2016characterization}
Henderson BS, Wernet M (2016) Characterization of three-stream jet flow fields.
  In: 54th AIAA Aerospace Sciences Meeting, p 1636

\bibitem[{Huff et~al.(2016)Huff, Henderson, Berton, and
  Seidel}]{huff2016perceived}
Huff DL, Henderson BS, Berton JJ, Seidel JA (2016) Perceived noise analysis for
  offset jets applied to commercial supersonic aircraft. In: 54th AIAA
  Aerospace Sciences Meeting, p 1635

\bibitem[{Jordan and Colonius(2013)}]{jordan2013wave}
Jordan P, Colonius T (2013) Wave packets and turbulent jet noise. Annual Review
  of Fluid Mechanics 45:173--195

\bibitem[{Keefe(2015)}]{keefe2015magazine}
Keefe S (2015) F/a-18 program explores the use of exhaust nozzle chevrons to
  reduce engine noise. Currents: The Navy's Energy and Environment Magazine pp
  6--19

\bibitem[{Lighthill(1952)}]{lighthill1952sound}
Lighthill MJ (1952) On sound generated aerodynamically i. general theory. Proc
  R Soc Lond A 211(1107):564--587

\bibitem[{Lilley(1996)}]{lilley1996radiated}
Lilley G (1996) The radiated noise from isotropic turbulence with applications
  to the theory of jet noise. Journal of sound and vibration 190(3):463--476

\bibitem[{Mayo et~al.(2017)Mayo, Daniel, Lowe, and Ng}]{mayo2017experimental}
Mayo DE, Daniel K, Lowe KT, Ng WF (2017) Experimental investigation of a heated
  supersonic jet with total temperature non-uniformity. In: 23rd AIAA/CEAS
  Aeroacoustics Conference, p 3521

\bibitem[{Mayo et~al.(2018)Mayo, Daniel, Lowe, and Ng}]{mayo2018flow}
Mayo DE, Daniel K, Lowe T, Ng W (2018) The flow and turbulence characteristics
  of a heated supersonic jet with an offset total temperature non-uniformity.
  In: 2018 AIAA/CEAS Aeroacoustics Conference, p 3144

\bibitem[{Morris and Zaman(2010)}]{morris2010velocity}
Morris PJ, Zaman KB (2010) Velocity measurements in jets with application to
  noise source modeling. Journal of sound and vibration 329(4):394--414

\bibitem[{Panda et~al.(2005)Panda, Seasholtz, and
  Elam}]{panda2005investigation}
Panda J, Seasholtz RG, Elam KA (2005) Investigation of noise sources in
  high-speed jets via correlation measurements. Journal of Fluid Mechanics
  537:349--385

\bibitem[{Papamoschou(2018)}]{papamoschou2018modelling}
Papamoschou D (2018) Modelling of noise reduction in complex multistream jets.
  Journal of Fluid Mechanics 834:555--599

\bibitem[{Papamoschou and Phong(2017)}]{papamoschou2017very}
Papamoschou D, Phong VC (2017) The very near pressure field of single-and
  multi-stream jets. In: 55th AIAA Aerospace Sciences Meeting, p 0230

\bibitem[{Papamoschou and Rostamimonjezi(2012)}]{papamoschou2012modeling}
Papamoschou D, Rostamimonjezi S (2012) Modeling of noise reduction for
  turbulent jets with induced asymmetry. In: 18th AIAA/CEAS Aeroacoustics
  Conference (33rd AIAA Aeroacoustics Conference), p 2158

\bibitem[{Papamoschou et~al.(2010)Papamoschou, Morris, and
  McLaughlin}]{papamoschou2010beamformed}
Papamoschou D, Morris PJ, McLaughlin DK (2010) Beamformed flow-acoustic
  correlations in a supersonic jet. AIAA journal 48(10):2445--2453

\bibitem[{Papamoschou et~al.(2014{\natexlab{a}})Papamoschou, Johnson, and
  Phong}]{papamoschou2014aeroacoustics}
Papamoschou D, Johnson AD, Phong V (2014{\natexlab{a}}) Aeroacoustics of
  three-stream high-speed jets from coaxial and asymmetric nozzles. Journal of
  Propulsion and power 30(4):1055--1069

\bibitem[{Papamoschou et~al.(2014{\natexlab{b}})Papamoschou, Xiong, and
  Liu}]{papamoschou2014reduction}
Papamoschou D, Xiong J, Liu F (2014{\natexlab{b}}) Reduction of radiation
  efficiency in high-speed jets. In: 20th AIAA/CEAS Aeroacoustics Conference, p
  2619

\bibitem[{Petitjean et~al.(2007)Petitjean, Viswanathan, McLaughlin, and
  Morris}]{petitjean2007space}
Petitjean B, Viswanathan K, McLaughlin D, Morris P (2007) Space-time
  correlation measurements in subsonic and supersonic jets using optical
  deflectometry. In: 13th AIAA/CEAS Aeroacoustics Conference (28th AIAA
  Aeroacoustics Conference), p 3613

\bibitem[{Shea et~al.(2017)Shea, Lowe, and Ng}]{shea2017eddy}
Shea S, Lowe KT, Ng WF (2017) Eddy convection in cold and heated supersonic
  jets. In: 23rd AIAA/CEAS Aeroacoustics Conference, p 4044

\bibitem[{Stuber(2017)}]{stuber2017investigation}
Stuber MA (2017) Investigation of noise sources in three-stream jets using
  turbulence characteristics. PhD thesis, Virginia Tech

\bibitem[{Unnikrishnan and Gaitonde(2017)}]{unnikrishnan2017acoustic}
Unnikrishnan S, Gaitonde DV (2017) Acoustic mode and sources in a supersonic
  jet. In: 55th AIAA Aerospace Sciences Meeting, p 0685

\bibitem[{Veltin et~al.(2011)Veltin, Day, and
  McLaughlin}]{veltin2011correlation}
Veltin J, Day BJ, McLaughlin DK (2011) Correlation of flowfield and acoustic
  field measurements in high-speed jets. AIAA journal 49(1):150--163

\bibitem[{Wernet and Wernet(1994)}]{wernet1994stabilized}
Wernet JH, Wernet MP (1994) Stabilized alumina/ethanol colloidal dispersion for
  seeding high temperature air flows. In: Proceedings of the ASME Symposium on
  Laser Anemometry: Advances and Applications

\bibitem[{Williams and Maidanik(1965)}]{williams1965mach}
Williams JF, Maidanik G (1965) The mach wave field radiated by supersonic
  turbulent shear flows. Journal of Fluid Mechanics 21(4):641--657

\end{thebibliography}

\end{document}